\documentclass[onecolumn,           
               noshowpacs,          
               nopreprintnumbers,     
               aps,                 
               prd,                 
               superscriptaddress,  
               nofootinbib,         
               tightenlines,        
               floats,floatfix ,     
               showkeys]{revtex4-2}

\usepackage{bm}
\usepackage{amsmath, amsfonts, amsthm, amssymb, graphicx}
\usepackage{epstopdf}
\usepackage{slashed}
 \usepackage{graphicx,epsfig}
\usepackage{amsmath}
\usepackage {amssymb}
\usepackage[utf8]{inputenc}
\usepackage{hyperref}
\usepackage{xcolor}
\usepackage[T1]{fontenc}
\usepackage{varwidth}
\usepackage{graphicx}
\usepackage{subcaption}

\begin{document}

\title{Teaching physics in real-life contexts: The Beirut explosion}

\author{Mauricio Echiburu}
\email{mechiburu@uvm.cl}
\affiliation{Escuela de Ciencias, Universidad Viña del Mar, Agua Santa 7055, Viña del Mar, Chile.}

\author{Carla Hern\'andez}
\email{carla.hernandez.s@usach.cl}
\affiliation{Departamento de F\'{i}sica, Universidad de Santiago de Chile, Avenida Víctor Jara 3493, Estaci\'on Central, Santiago, Chile.}
\affiliation{Center for Interdisciplinary Research in Astrophysics and Space Exploration (CIRAS), Universidad de Santiago de Chile, Avenida Libertador Bernardo O'Higgins 3363, Estación Central, Chile.}
 
\author{Miguel Pino}
\email{miguel.pino.r@usach.cl}
\affiliation{Departamento de F\'{i}sica, Universidad de Santiago de Chile, Avenida Víctor Jara 3493, Estaci\'on Central, Santiago, Chile.}
\affiliation{Center for Interdisciplinary Research in Astrophysics and Space Exploration (CIRAS), Universidad de Santiago de Chile, Avenida Libertador Bernardo O'Higgins 3363, Estación Central, Chile.}


\begin{abstract}

  Teaching physics in real-life contexts continues to be a challenge for teachers at different educational levels. In this article, three Context-Rich Problems are proposed to be implemented in the classroom for higher education, using the explosion that occurred in Beirut as a case study. These problems require the search and analysis of real data, integrating technologies as tools to learn physics. In particular, the analysis of images, videos, maps and audio recordings is suggested. The proposed activities are designed to promote active learning of classical mechanics subjects and the development of collaborative skills. The results of each problem allow a discussion based on scientific evidence in the classroom.

\end{abstract}


\maketitle
\section{Introduction}

On August 4, 2020, a big explosion occurred in the city of Beirut, Lebanon, killing 218 people and injuring about 7000, generating great expectation and concern worldwide.  In order to understand the causes, the impact and the prevention of similar tragedies, it is required that citizens carry out a critical analysis of the situation. The promotion of scientific literacy among citizens has been considered in recent decades as one of the main objectives of science education in the study programs of all countries \cite{she, kembara}. In coherence, for this article we consider that the explosion represents a valuable educational opportunity to promote teaching of physics applied to real contexts, based on the critical analysis of the information obtained through the media. 

Problem Solving is one of the main activities carried out in physics classes, which allow students to develop important skills \cite{Mestre,OECD,Larkin,Leonard}. When faced with a problematic situation, a person needs to make decisions in order to advance towards a goal without knowing the path \cite{Martinez}. However, it is common for traditional physics classes to pose problems in ideal contexts, where the connections between concepts, formal representations and reality are practically non-existent \cite{Hernandez}. Furthermore, these activities are generally oriented to the application of a specific content or to the calculation of certain variables, which requires the use of a single physical principle or an established procedure for their resolution \cite{Dufresne}.

Given the above, some strategies have been developed in recent decades which promote Problem Solving in the classroom by means of an analysis similar to those carried out by experts, on the basis of logically organized steps, and allowing the student to understand what they are doing \cite{Heller_0}. Cooperative problem solving, through the so-called Context-Rich Problems (CRP) \cite{Heller}, is a good active learning strategy that allows to promote qualitative and quantitative reasoning, the prediction of results and their verification, as well as the integration of knowledge. These problems are designed to be solved by more than one person and are presented as a story regarding a real situation.

As shown in the activities below, the Beirut explosion is an opportunity to motivate students to apply physical concepts within such a context, while promoting group work due to its complexity \cite{Benegas}. According to the recommendations found in \cite{Heller}, we can encourage students to adopt specific roles, for example, as researchers or specialists making decisions based on evidence, performing qualitative estimates and analyzing the situation. In this way, students develop skills such as synthesis capacity, communication of ideas, construction of models and the interpretation of phenomena.

Unlike a traditional problem, where the necessary data is informed in the statement, obtaining such information in a CRP can also be part of its resolution. Given the overcrowding of technology and its emergence in the classroom for teaching and learning at the end of the 20th century, several techniques that encourage the analysis of data from videos and images of real phenomena have been promoted. The first reports on video analysis in physics classes suggested a positive effect on student attitudes \cite{Escalada,Lewis}. Video analysis has been widely used and documented, standing out for its economy and for allowing students to make accurate measurements of various phenomena involving moving objects \cite{Bryan,Phommarach,Vera,Grober}. Several tools are available for this purpose, such as PhysTrack from Matlab \cite{Hassan} or Tracker software \cite{Wee,MuliyatiA}, which is free to access online.

The Beirut explosion has been widely analyzed in the literature by means of high-end image processing techniques \cite{Diaz,Dewey,Aouad,Rigby}. In this paper we present three Context-Rich Problems aiming to perform similar scientific analysis in the classroom for university level. The solution of these activities entails basic physical concepts and the use of easily attainable tools. In particular, these proposals promote active learning of the following topics: dimensional analysis, the energy released by the explosion and wave propagation.

Each of the next three sections presents a self contained Context-Rich Problem. Within them there is a brief introduction, an explicit formulation of the problem, written to be presented to the students, followed by a proposed solution along with the respective data acquisition. Ending remarks can be found at the end of the paper.

\section{Problem 1: Dimensional analysis of an explosion}

The overwhelming availability of videos recorded during the Beirut explosion yields a motivation to critically analyze this situation from a scientific perspective in the classroom . In particular, several videos capture the precise moment of the explosion (for example, some of the videos in Table \ref{datos}), where a spherical fireball can be observed violently increasing its radius. In order to quantitatively analyze these recordings, we first need to construct a suitable dynamical model for the fireball expansion, a task which is out of the scope of a beginner course on physics \cite{Taylor 1, Taylor 2}. However, it can be drastically simplified by means of dimensional analysis. This CRP deals with such construction.  

\subsection*{Formulation of the problem}

\textit{Your work team is requested to write a technical report regarding an explosion occurred in an industry. In the attached documentation you find a video footage of the explosion site, where a spherical fireball can be observed increasing its radius in time. Upon discussion with your team, you agree that it is reasonable to assume that the precise form on which the radius of the fireball $R$ changes in time $t$ depends on the energy released in the explosion $E$ and on the density of the surrounding air $\rho$. You aim to construct the mathematical model for such a process. A colleague on your team remembers that the model can be determined by using dimensional analysis. At the end of the day your team is satisfied with the work done and they receive congratulations from the management.  }
\subsection*{Proposed solution}

Assuming that the explosion occurs with spherical symmetry, the radius $R$ of the fireball should depend on the energy released in the detonation $E$, on the air density $\rho$, and on time $t$. A suitable ansatz is given by
\begin{equation} \label{eq2}
R=S E^a\rho^b t^c,
\end{equation}
where $S$ is a dimensionless factor, while $a$, $b$ and $c$ are real numbers to be determined by dimensional analysis (see \cite{Bridgman} for a review). Let $L$ stand for the dimension of length, $M$ for the dimension of mass and $T$ for the dimension of time. Hence, equation \eqref{eq2} yields
\begin{equation} \label{eq3}
    L=(ML^2T^{-2})^a (ML^{-3})^b T^c,
\end{equation}
or equivalently, 
\begin{equation}\label{eq4}
    L=M^{a+b}L^{2a-3b}T^{c-2a}.
\end{equation}
The consistency of the previous relation implies
\begin{align*} 
a+b=0, \\
2a-3b=1,\\
  -2a+c=0,
\end{align*}
\vspace{10pt}
whose solution is $a = 1/5$, $b = -1/5$ and $c = 2/5$. Accordingly, equation \eqref{eq2} becomes 
\begin{equation} \label{eq5}
R=S E^{1/5}\rho^{-1/5} t^{2/5},
\end{equation}
which is the sought model.

It is noteworthy that the above model \eqref{eq5}, obtained using very simple arguments, coincides with the detailed analysis carried out by Sir Geoffrey Taylor published in 1950 \cite{Taylor 1, Taylor 2} in the context of nuclear explosions.

As a final comment for this section, it is worth mentioning that the presented solution does not involve the dimensionless constant $S$ in any way, ignoring its physical meaning. As in any dimensional analysis problem, in absence of further information this constant is assumed to be of order one, as it will be the case in the following problem. However, in the aforementioned publications \cite{Taylor 1, Taylor 2}, it is shown that $S$ is actually a function of the ratio of the specific heats of air, $\gamma$. Such a result might be fruitful to mention to the students, but otherwise completely out of the scope of this problem.

\section{Problem 2: Energy released in an explosion from video analysis}

Having obtained the model of the previous problem, a quantitative analysis of real data obtained from the video recordings of the Beirut explosion can be performed. In particular, this problem will concern the YouTube video \cite{video}, where the explosion can be clearly observed and measured.      

In order to make this CRP self contained, the model \eqref{eq5} is included in the formulation of the problem, expressed in the form  
\begin{equation}\label{eq6}
    \frac{5}{2}\log (R)=\frac{5}{2}\log(S E^{1/5} \rho^{-1/5})+\log(t).
  \end{equation}
  
Values for the fireball radius $R$ at different times $t$ can be obtained from analyzing the video recording using suitable software. Considering equation \eqref{eq6}, and assuming adequate values for the air density $\rho$ and the dimensionless constant $S$, a linear fit can be used to estimate the energy $E$ released in the explosion. For this objective, the following CRP is proposed, on which the students are encouraged to perform the data acquisition.

\subsection*{Formulation of the problem}

\textit{Through social media, you heard about a great and impressive accidental explosion. The press indicates that the energy released in the accident is similar to an atomic bomb. You are doubtful about such a claim and therefore you want to verify it. On the internet you find a video (\url{https://youtu.be/4DH0URNodcQ}) where you can observe the explosion fireball increasing its radius in time. Along with your friends, you find a relation among the fireball radius $R$, the energy released $E$, the density of the air $\rho$, and time $t$, given by: $\frac{5}{2}\log(R)=\frac{5}{2}\log(S E^{1/5}\rho^{-1 / 5}) + \log (t) $. Here, $S$ is a dimensionless coefficient whose value is approximately 1. From the video you can measure the fireball radius at different times from a sequence of frames, allowing you to find the energy released in the explosion.}

\subsection*{Proposed solution}

Regarding the variables involved in the problem, air density is the simplest to deal with; it can be easily found online, roughly $ \rho=1.17\ \mathrm{kg/m^3}$. Minor differences due to temperature and humidity will not result in a substantial variation of the final result.

The fireball radius $R$ increases with time according to the equation presented in the problem; it corresponds to a straight line whose abscissa is $\frac{5}{2} \log (R)$ and ordinate $\log (t)$. The value of the intercept $b$ is equal to
\begin{equation}\label{eq7}
    \frac{5}{2}\log(S E^{1/5}\rho^{-1/5}). 
\end{equation}
Hence, the energy released in the explosion can be determined as
\begin{equation}\label{eq8}
    E=\frac{\rho}{S^5} 10^{2b}.
\end{equation}

\begin{figure}[htb]
  \centering
  \begin{subfigure}[b]{.19\linewidth}
    \centering
    \includegraphics[width=.99\textwidth]{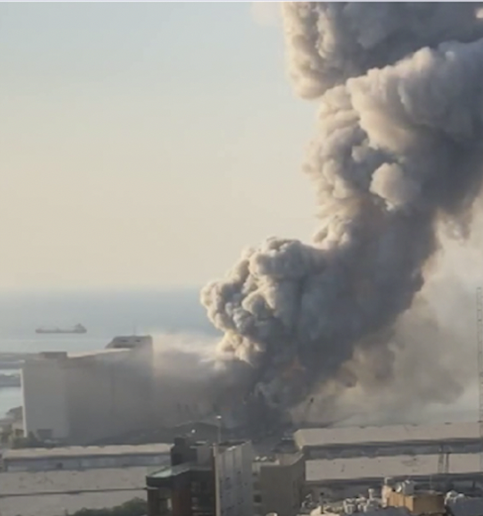}
    \vspace{-15pt}
    \caption*{$t_{-1}=-33.3$ ms}
  \end{subfigure}%
  \begin{subfigure}[b]{.19\linewidth}
    \centering
    \includegraphics[width=.99\textwidth]{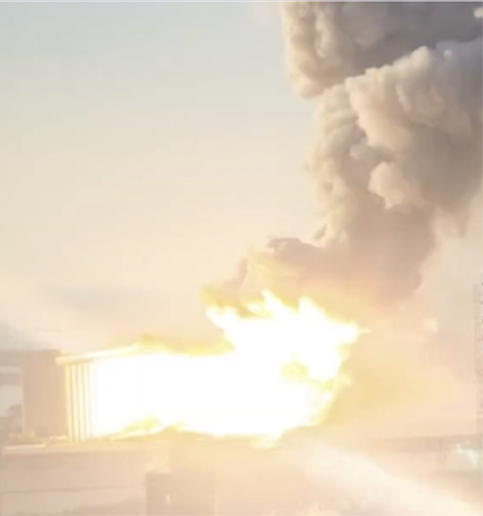}
    \vspace{-15pt}
    \caption*{$t_{0}=0.0$ ms}
  \end{subfigure}%
  \begin{subfigure}[b]{.19\linewidth}
    \centering
    \includegraphics[width=.99\textwidth]{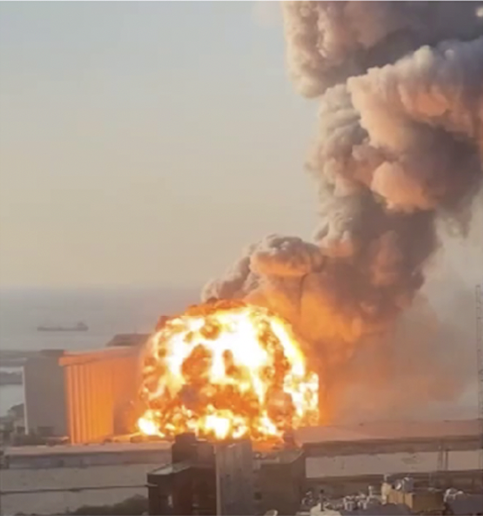}
    \vspace{-15pt}
    \caption*{$t_{1}=33.3$ ms}
  \end{subfigure}%
  \begin{subfigure}[b]{.19\linewidth}
    \centering
    \includegraphics[width=.99\textwidth]{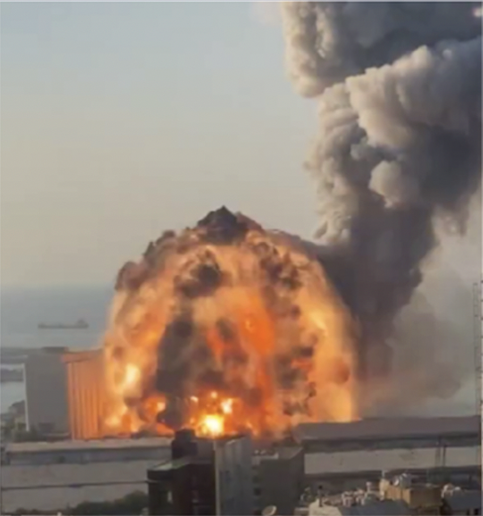}
    \vspace{-15pt}
    \caption*{$t_{2}=66.6$ ms}
  \end{subfigure}%
  \begin{subfigure}[b]{.19\linewidth}
    \centering
    \includegraphics[width=0.99\textwidth]{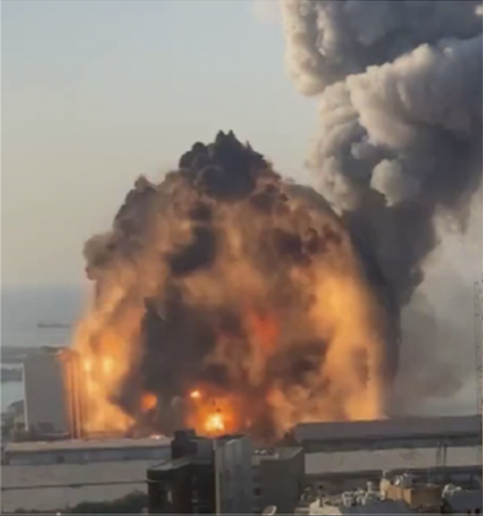}
    \vspace{-15pt}
    \caption*{$t_{3}=99.9$ ms}
  \end{subfigure}
  \vspace{5pt}
  \begin{subfigure}[b]{.19\linewidth}
    \centering
    \includegraphics[width=.99\textwidth]{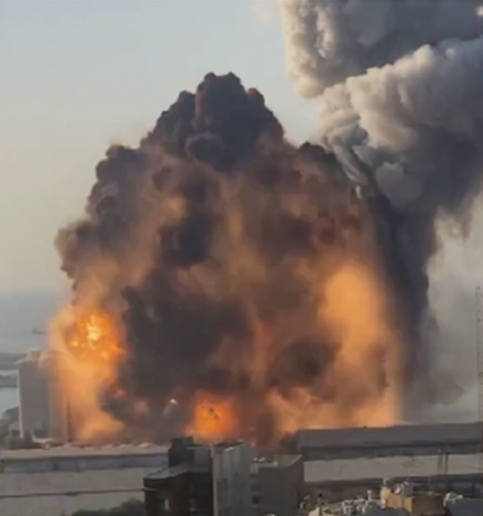}
    \vspace{-15pt}
    \caption*{$t_{4}=133.2$ ms}
  \end{subfigure}%
  \begin{subfigure}[b]{.19\linewidth}
    \centering
    \includegraphics[width=.99\textwidth]{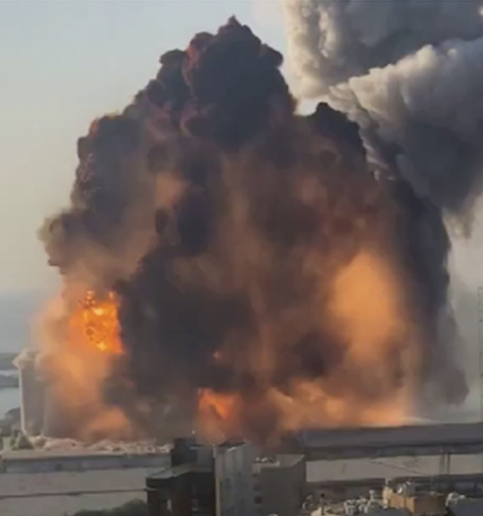}
    \vspace{-15pt}
    \caption*{$t_{5}=166.5$ ms}
  \end{subfigure}%
  \begin{subfigure}[b]{.19\linewidth}
    \centering
    \includegraphics[width=.99\textwidth]{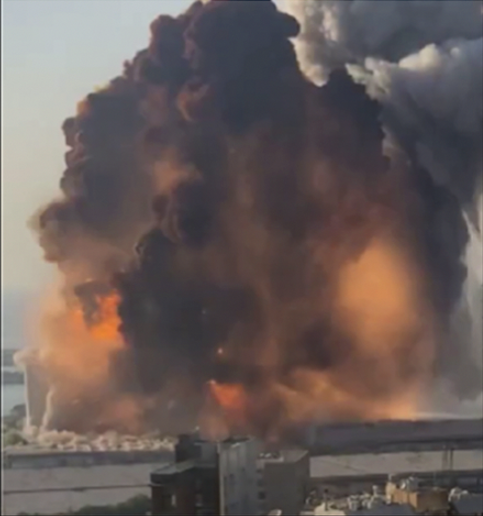}
    \vspace{-15pt}
    \caption*{$t_{6}=199.8$ ms}
  \end{subfigure}%
  \begin{subfigure}[b]{.19\linewidth}
    \centering
    \includegraphics[width=.99\textwidth]{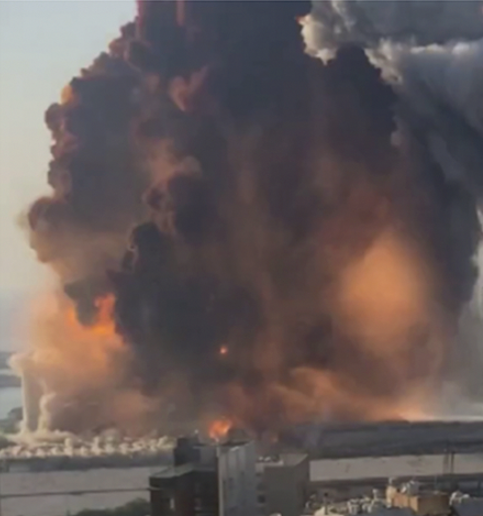}
    \vspace{-15pt}
    \caption*{$t_{7}=233.1$ ms}
  \end{subfigure}%
  \begin{subfigure}[b]{.19\linewidth}
    \centering
    \includegraphics[width=0.99\textwidth]{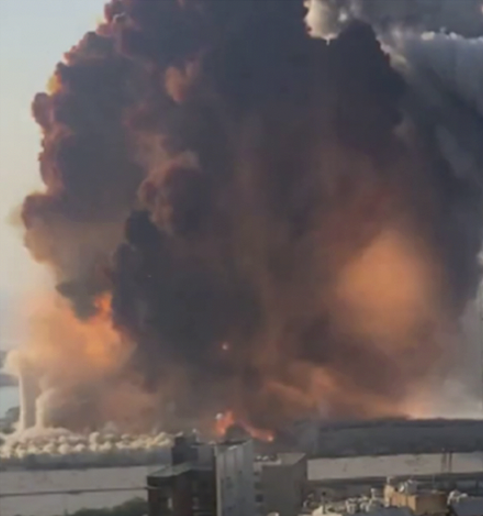}
    \vspace{-15pt}
    \caption*{$t_{8}=266.4$ ms}
  \end{subfigure}%
  \caption{Image sequence of the 300 ms explosion. The building on the left of the fire in frame $t_{-1}$ is used as a reference width.}
  \label{colach}
\end{figure}

The value of the intercept $b$ can be found from the analysis of the indicated video, which was recorded by Agoston Nemeth and published by CNN \cite{video}. For this purpose, it is possible to measure $R$ and $t$ from a sequence of photographs obtained using Tracker\footnote{\url{https://physlets.org/tracker/}}. This software works at 30 frames per second, which corresponds to a time between frames of $33.3$ ms, as shown in Figure \ref{colach}.

To measure the radius $R$ of the fireball, it is necessary to identify a reference length in the images. The building right next to the shed where the fire started can be used in this regard, since it can be easily spotted in the video and its width can be measured using satellite pictures, such as Google Earth. The apparent width of the building in the video is obtained by identifying the position and orientation of Agoston's home with respect to the explosion site, which can be accomplished in Google Earth by carefully observing the nearby buildings, as shown in Figure \ref{referencias}.

\begin{minipage}{\linewidth}
  \vspace{10pt}
  \begin{minipage}{0.5\linewidth}
    \centering
    \includegraphics[width=0.8\columnwidth]{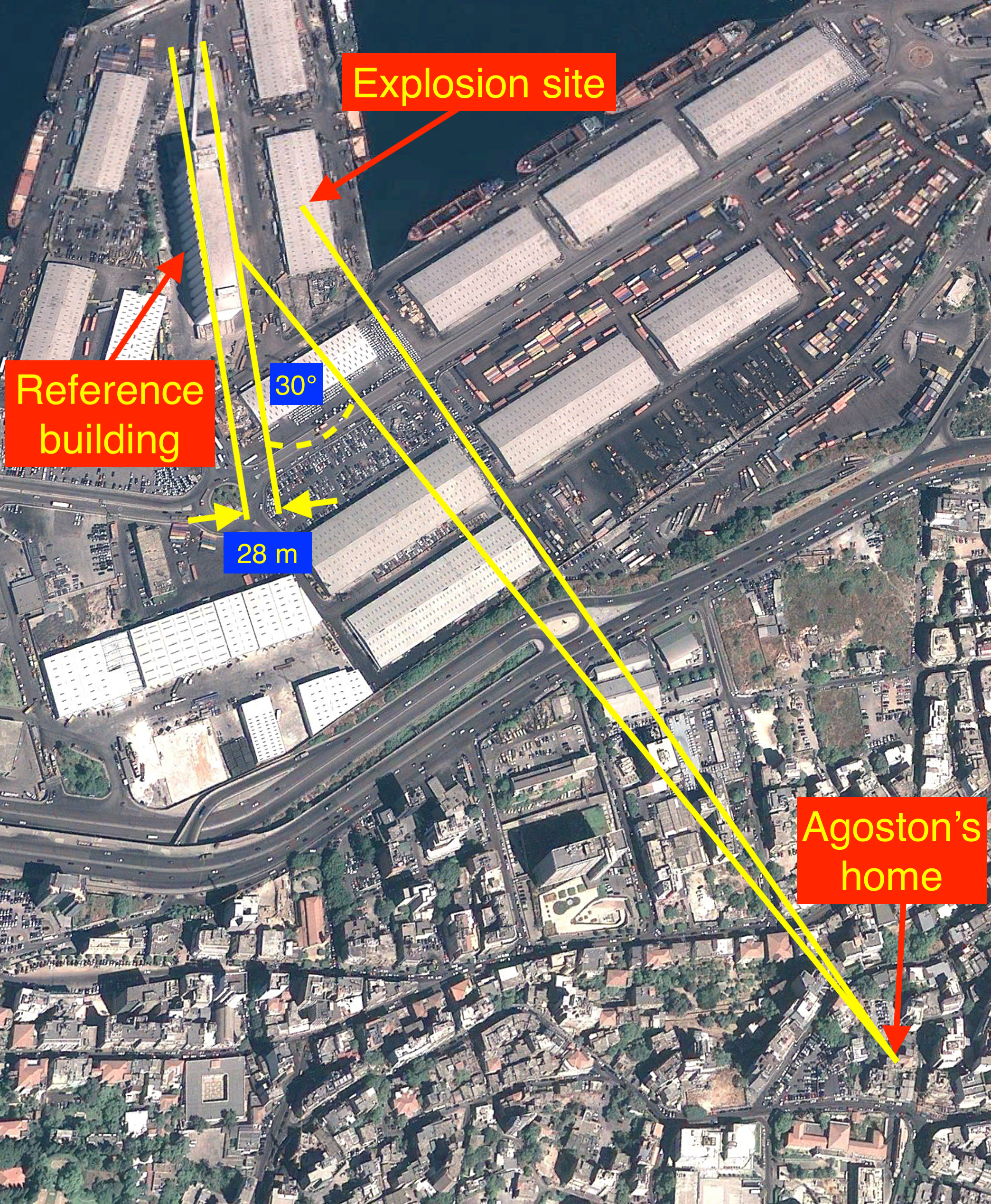}
    \captionof{figure}{Map showing the location of Agoston Nemeth's home and the site of the explosion. The reference building can be clearly seen on frame $t_{-1}$ of Figure \ref{colach}. Source: Google Earth.}
    \label{referencias}
  \end{minipage}  
  \begin{minipage}{0.5\linewidth}
    \vspace{20pt}
    \centering    
    \begin{tabular}{c c c c} 
      \hline
      \textbf{$\bm{R}$ (m)} & \textbf{$\bm{t}$ (ms)} & $\bm{\frac{5}{2} \log  R}$ & $\bm{\log t}$ \vspace{1pt}\\ 
      \hline
      52 &   33.3 & 4.28 & -1.49 \\
      
      74 & 66.6 & 4.67 & -1.18\\
      
      99 & 99.9 & 4.99 & -1.00\\
      
      108 & 133.2 & 5.08 & -0.87\\
      
      119 & 166.5 & 5.18 & -0.78\\ 
      \hline
    \end{tabular}
        \vspace{30pt}
    \captionof{table}{Fireball radius $R$ (in meters) and frame time (in milliseconds).}
    \label{table1}
  \end{minipage}
\end{minipage}
\\


      With the considerations mentioned above, and using Google Earth distance tool, the apparent width of the reference building is roughly 24 m. Consequently, the data in Table \ref{table1} can be obtained from the photograph sequence.

      \begin{figure}[h!] 
        \centering \includegraphics[width=1\columnwidth]{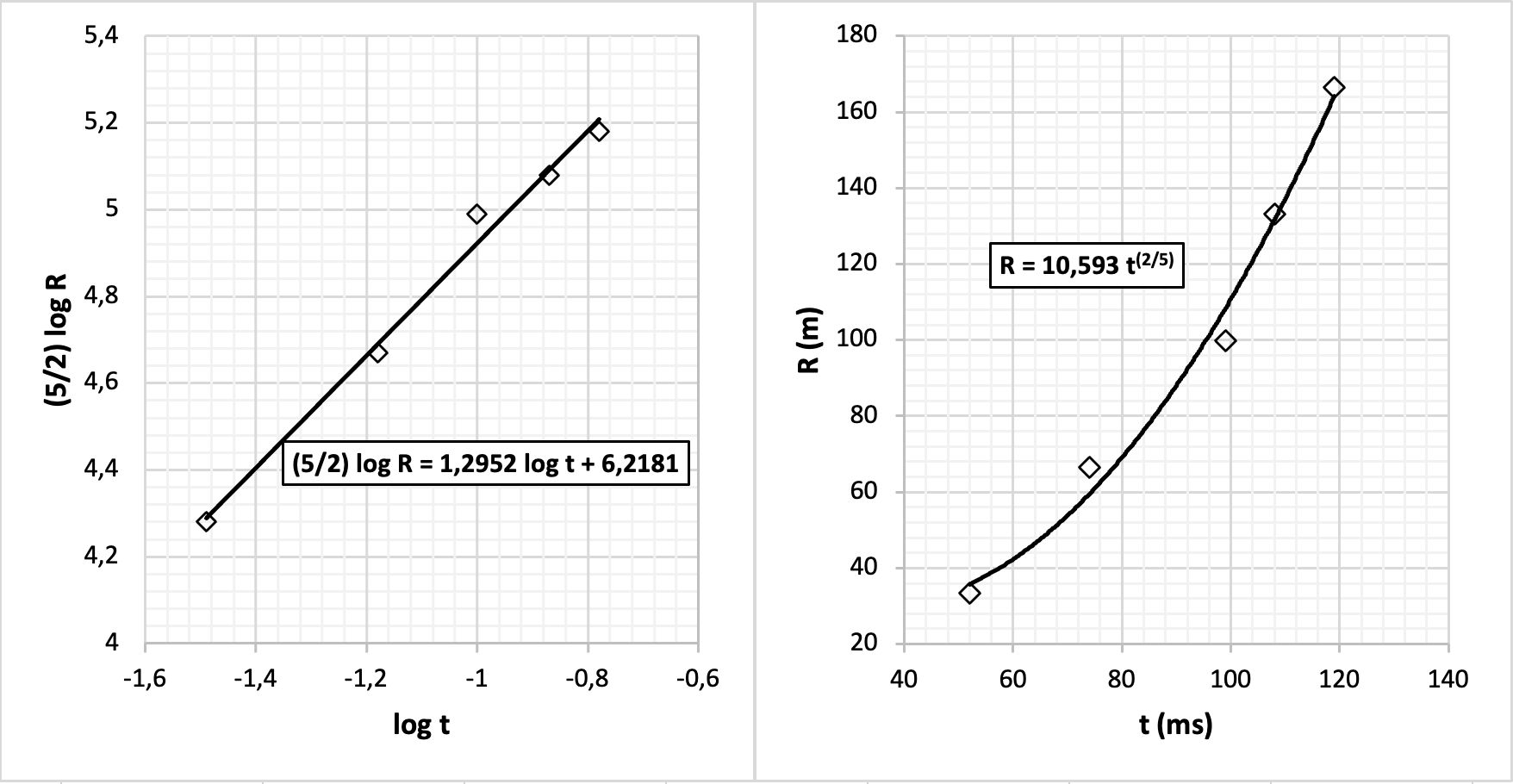}
        \caption{Left: Linear regression of data from Table \ref{table1}. Right: Power law fit of the same data.}
        \label{plots}
      \end{figure}

      
      By performing a linear fit, the intercept $b$ can be obtained, as shown by the plot at the left of Figure \ref{plots}. Hence, considering equation \eqref{eq8}, the value of the energy is then given by $E=4.25\times 10^{12}$ J.
      
The equation presented in the formulation of the problem is designed as a suggestion to the students to implement the aforementioned linear fit. This method is simple enough to obtain a good estimation of the intercept $b$ even by pencil and ruler methods. Alternatively, if the use of computers and software is available in the classroom, a more sophisticated fit can be easily performed using, for example, Python. The plot at the right of Figure \ref{plots} corresponds to the latter procedure. If such an approach is chosen, the teacher should modify the formulation of the problem, including equation \eqref{eq5} instead of its logarithmic version.

As a way to improve the discussion among students, the teacher can establish some comparisons. For example, considering that 1 kiloton (1 kt) is equal to $ 4.184 \times 10^{12}$ J, the result obtained here corresponds roughly to 1 kt. This coincides with the press reports that indicated an energy release of 1 to 3 kt. For comparison, the Hiroshima bomb "Little Boy" released 16 kt in the explosion.

As explained in the preceding problem, the dimensionless constant $S$, which is assumed to be approximately 1, might differ from such a value. The teacher should encourage the search in \cite{Taylor 2} for more realistic values for $S$, motivating the discussion on how such variation affects the obtained value for the energy, and how this source of error compares to those coming from the measurements.

It is worth mentioning that for the purposes of this problem, other videos available in social media can be used. The necessary condition is that a length reference can be identified in the images. Furthermore, similar analysis has been carried out in \cite{Diaz,Dewey,Aouad,Rigby}, using far more sophisticated image processing tools.

Finally, it should be emphasized that with this problem the students will be able to learn and practice topics such as unit conversion, construct logarithmic graphs and the use of software useful for physics.

\section {Problem 3: Wavefront velocities from video recordings}

Among the videos recorded during Beirut's explosion, several of them do not show the fireball itself, but the reaction of people in nearby places. As it can be observed on the YouTube videos of Table \ref{datos}, on each location two clearly noticeable moments can be identified. First, a comparatively low intensity sound wave reaches the observers (front A), preceded by a high intensity shock wave (front B). However, the time interval between the arrival of both fronts is different for each location, indicating that the wavefronts propagate with different velocities.

Additional information can be obtained from the videos whose location can be pinpointed; using Google Earth distance tool, the distance from such a places to the explosion site can be measured. For those videos whose location cannot be identified, the distance to the explosion cannot be directly obtained.   

The following problem is designed to analyze the propagation velocity of each front by measuring the time of arrival differences on each video. The purpose is to establish a procedure to estimate the distance between the explosion site and the recording location of those videos that cannot be placed.

It is worth stressing the importance of internet availability and a suitable sound system in the classroom for this problem, in order to aid the explanation of the situation and to assure that the difference between each video is clear. The videos shown in Table \ref{datos} are preselected for the purposes of this problem.

\subsection*{Formulation of the problem}

\textit{You are reviewing internet videos of the Beirut explosion occurred in 2020. You realize that the time elapsed between the arrival of the explosion sound (front A) and the arrival of the shock wave (front B) is different on each recording, suggesting that these two wavefronts propagate with different velocities. Furthermore, the location where some of the videos where filmed can be identified, and you can easily determine how far away the explosion occurred using satellite images. Given your interest in physics, you decide to design a procedure to estimate how far away any camera that recorded the event was from the explosion. }

\subsection*{Proposed solution}

It is assumed that both fronts are produced simultaneously at the explosion site. In what follows, $D$ represents the distance between the recording location and the explosion, while $\Delta t$ is the time interval between the arrival of both fronts. It is straightforward to show that the following relation holds
\begin{equation}\label{ecc}
  \Delta t = D \left( \frac{1}{v_B}-\frac{1}{v_A} \right),
\end{equation}
where $v_A$ and $v_B$ stand for the mean propagation velocity of each front. The values of $D$ and $\Delta t$ can be measured from each video, as explained below. A linear fit of the data obtained from the videos can then be used to get the value of the difference $\frac{1}{v_B}-\frac{1}{v_A}$.

\begin{figure}[h!]
\includegraphics[width=1\textwidth]{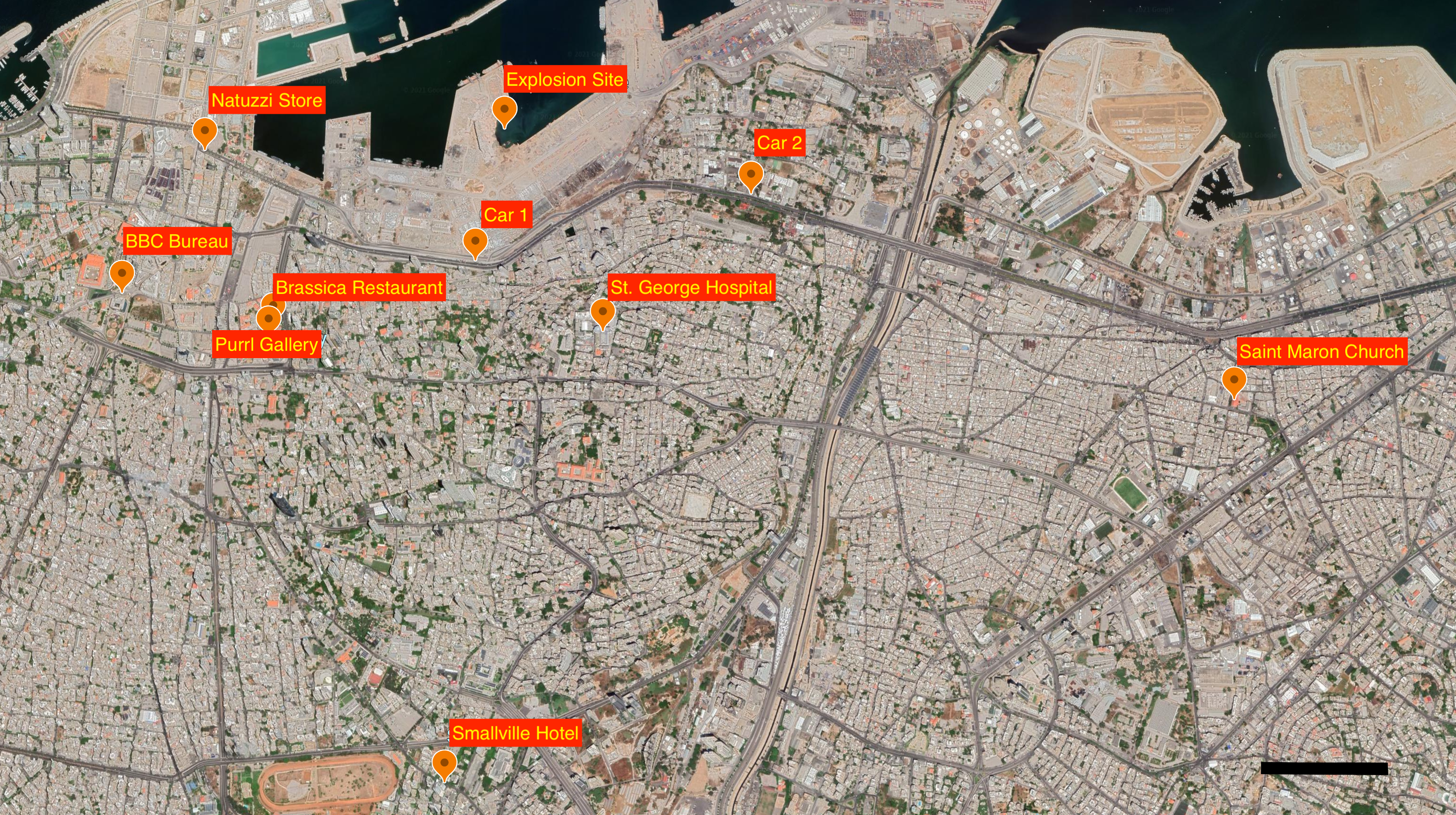}
  \caption{Google Earth view of downtown Beirut, indicating the explosion site and the location of some video recordings. The black strip in the lower right corner corresponds to a distance of 500 m.}
  \centering
  \label{mapa}
\end{figure}

To measure the distance $D$ on each video, first the location of the recording must be identified by observing features in the images, such as store's names, nearby buildings, additional information included in the video, etc. Once a successful identification of the location is accomplished, the distance $D$ to the explosion site can be determined using the rule tool in Google Earth. The map in Figure \ref{mapa} shows the identified location of some videos. Several other videos do not have enough information to identify the precise location of the recording. Table \ref{datos} summarizes the location information.

\begin{table}[b]
\centering
\begin{tabular}{@{\extracolsep{\fill}}llrr}
  \hline 
  \textbf{Video link} & \textbf{Location} & \textbf{ $\bm{D}$ (km)} & \textbf{ $\bm{\Delta t}$ (s)}  \tabularnewline
                                                                       \hline
\href{https://youtu.be/_L7SlqDtRnc}{https://youtu.be/\_L7SlqDtRnc}&	Purrl Gallery&	1.2&	2.5\tabularnewline
\href{https://youtu.be/VBV8nnp98vc}{https://youtu.be/VBV8nnp98vc}&	Brassica Restaurant&	1.3&	2.5\tabularnewline
\href{https://youtu.be/2kXGrv8IRXU}{https://youtu.be/2kXGrv8IRXU}&	Saint Maron Church, Baouchrieh&	3.2&	7.1\tabularnewline
\href{https://youtu.be/SVpM2nQBFk8}{https://youtu.be/SVpM2nQBFk8}&	Natuzzi Store&	1.2&	2.1\tabularnewline
\href{https://youtu.be/yzBAChQu0co}{https://youtu.be/yzBAChQu0co}&	St. George Hospital (childbirth)&	1.0	&1.8\tabularnewline
\href{https://youtu.be/JIxuwE_WPXw}{https://youtu.be/JIxuwE\_WPXw}&	St. George Hospital (CCTV)&	1.0&	2.0\tabularnewline
\href{https://youtu.be/j1iwI2ZwENg}{https://youtu.be/j1iwI2ZwENg}&	BBC bureau&	1.7	&3.5\tabularnewline
\href{https://youtu.be/VQ61Lixmixg}{https://youtu.be/VQ61Lixmixg}&	Smallville Hotel&	2.7&	5.0\tabularnewline
\href{https://youtu.be/oFa6LV8EEZk}{https://youtu.be/oFa6LV8EEZk}&	Car 1, Charles Herou st. / Berberi st.&	0.6&	1.1\tabularnewline
\href{https://youtu.be/LFDj_2Lloks}{https://youtu.be/LFDj\_2Lloks}&	Car 2, Charles Herou st., between El Khodor st. and Ibrahim Bacha st.&	1.0&	1.9\tabularnewline
\href{https://youtu.be/uuduBHY_8Eo}{https://youtu.be/uuduBHY\_8Eo}&	Unidentified apartment 1&0.3*	&	0.6\tabularnewline
\href{https://youtu.be/2mBF2gSEEHQ}{https://youtu.be/2mBF2gSEEHQ}&	Unidentified apartment 2&2.4*	&	5.2\tabularnewline
\href{https://youtu.be/Vb8EXpZb7P4}{https://youtu.be/Vb8EXpZb7P4}&	Unidentified rooftop 1	&1.1*&	2.4\tabularnewline
\href{https://youtu.be/93tV6-0Ugwk}{https://youtu.be/93tV6-0Ugwk}&	Unidentified rooftop 2	&1.0*&	2.2\tabularnewline
\href{https://youtu.be/Nr9_kvw2aO0}{https://youtu.be/Nr9\_kvw2aO0}&	Unidentified street 	&0.5*&	1.0\tabularnewline
\href{https://youtu.be/8l4zl31QQRY}{https://youtu.be/8l4zl31QQRY}&	Unidentified church	&3.9*&	8.4\tabularnewline
\hline 
\end{tabular}
\caption{Identified location, distance to the explosion site $D$ and time interval between the arrival of both fronts $\Delta t$. The distance $D$ of those videos whose location cannot be identified, marked with an asterisk, is an estimate based on the linear fit of the available data.}
\label{datos}
\end{table}

\begin{figure}[h!]
\includegraphics[width=1\textwidth]{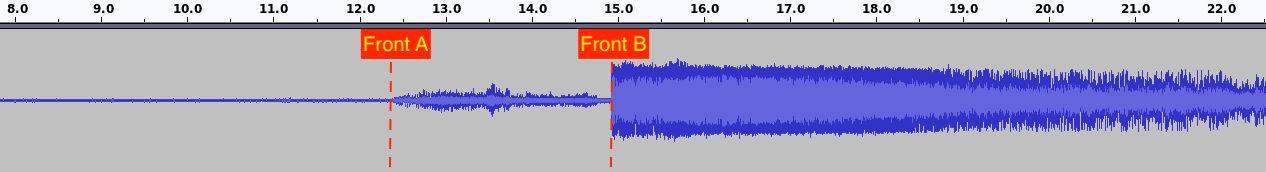}
  \caption{Audio channel of the video taken at Purrl Gallery (\href{https://youtu.be/_L7SlqDtRnc}{https://youtu.be/\_L7SlqDtRnc}), obtained using Audacity software (\url{https://www.audacityteam.org/}). The arrival of both fronts can be easily distinguished. The time scale corresponds to seconds.}
  \centering
  \label{audio}
\end{figure}

Regarding the measurement of $\Delta t$, for those videos with an available and low-noise recording, such interval can be measured by analyzing the video's audio track in a suitable software. Figure \ref{audio} shows an example of an audio track where the arrival of fronts A and B are clearly identifiable. If the audio track is absent or has a poor quality, the time interval can be measured by performing a frame by frame observation of the recording in a video software (similarly to problem 2). The measurements of $\Delta t$ are summarized in Table \ref{datos}.

\begin{figure}[h!]
  \centering
  \includegraphics[width=0.6\textwidth]{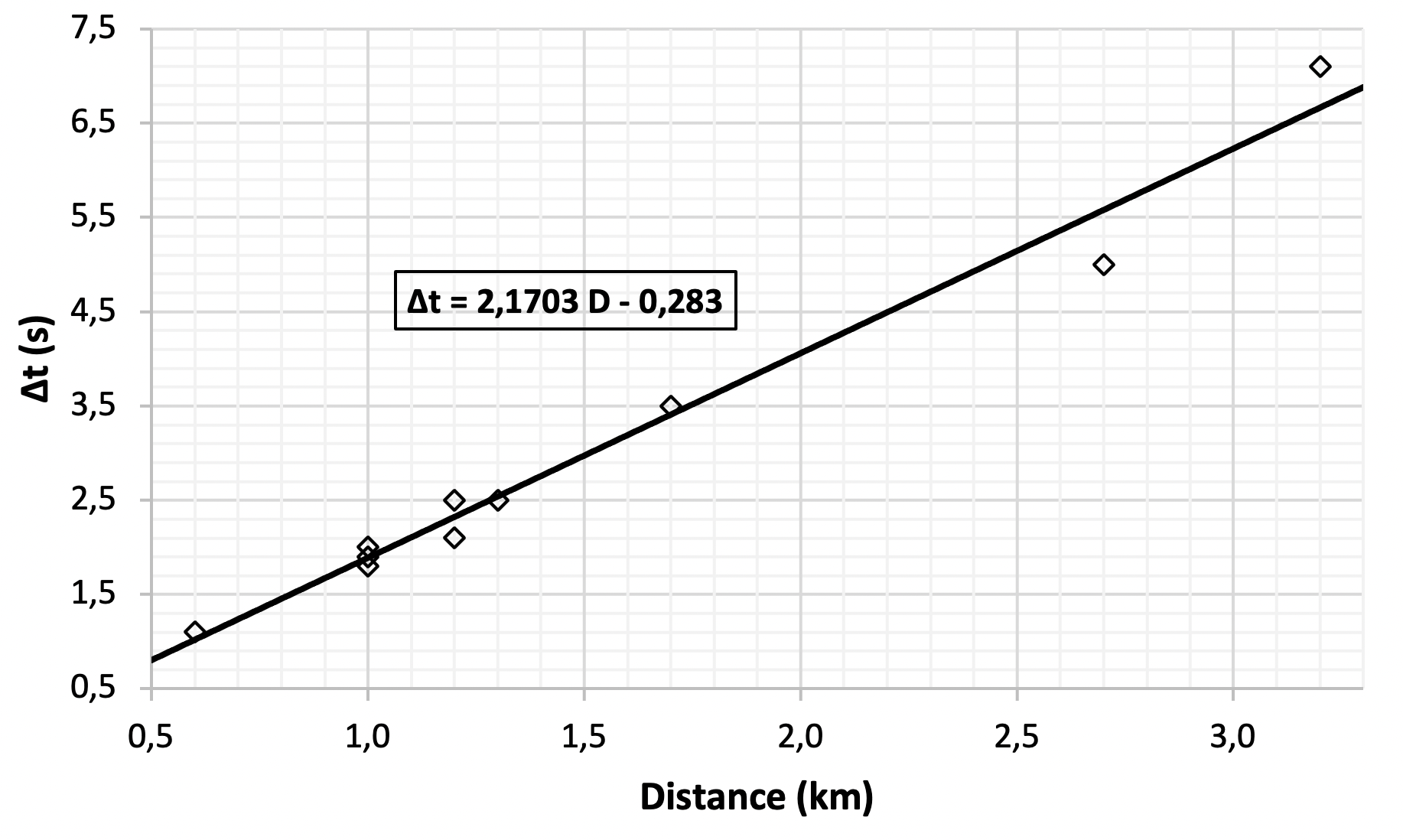}
  \caption{Linear fit of arrival time difference $\Delta t$ versus distance to the explosion site $D$.}
  \label{fit}
\end{figure}

Figure \ref{fit} shows the measured data of Table \ref{datos} with a linear fit. It yields a value $\left( \frac{1}{v_B} - \frac{1}{v_A}\right)\approx 2.2 \;\frac{\mbox{s}}{\mbox{km}} $. This can be used to get an estimate for the distances $D$ of the videos whose location identification is not possible. These estimations are indicated in Table \ref{datos} with an asterisk.

It is worth mentioning that this result assumes that the velocities $v_A$ and $v_B$ are constant along the propagation of the fronts. The linear fit seems to reassure this assumption, at least in the plotted distance range. The teacher should encourage the discussion of this matter, for example, by asking whether this condition should hold near the explosion or very far from it, or perhaps to decide if new data or observations are necessary to investigate this issue.

The analysis does not yield individual values for $v_A$ or $v_B$, but rather for the combination $\left( \frac{1}{v_B} - \frac{1}{v_A}\right)$. The teacher should use this observation to motivate the natural question regarding the nature of the fronts. Since it arrives earlier, front A is most likely a seismic wave propagating through the ground\footnote{The authors would like to thank the referee for pointing out this fact.}, while the shock wave front B propagates through the air. Although the objective of this problem is to discuss kinematics, this is a good opportunity to introduce the students to the propagation velocity of waves in different media.

As a final comment, if the students do not have access to suitable software or computers to carry the full analysis of the videos, the teacher can share the printed map \ref{mapa}  and audio tracks in order for the students to obtain the data using a ruler.

\section {Ending remarks}

The CRPs proposed in this work aim to promote collaborative and active learning in the classroom, as well as the development of problem solving and critical thinking skills. Each activity is formulated by presenting a real situation to the student, whose solution implies the design of a work scheme where scientific concepts should be applied \cite{Whitelegg}. Additional guidelines for the students regarding collaborative problem solving can be found in the literature \cite{Heller}.

Unlike traditional problems, the acquisition of data is part of the solution of the activities presented here. Consequently, the use of technologies is promoted, which implies access to computers and internet in the classroom. The use of free software is also encouraged. However, the unavailability of digital media should not be considered a restriction, since the teacher can adapt the problem by providing the necessary data to the students, such as the data tables, maps, images, etc., suggested in the solutions.

Appropriate planning is required for the teacher to allocate the necessary time to carry out these activities in the classroom. Solving a Context-Rich Problem such as those proposed here might require between 45 and 60 minutes if the data is delivered by the teacher, or about 90 minutes if the data acquisition is promoted as part of the resolution. Although each problem is independent and self-contained, these three activities can also be regarded as a pedagogical sequence.

Critical analysis of the obtained results as part of problem solving is a practice that should be strongly encouraged. For example, the results can be compared with the official information found in the media. In a globalized world, where information regarding events such as the Beirut explosion quickly floods the internet, it is easy for people to get a deformed opinion based on the lack of scientific education \cite{Howell,peters}. The classroom stands as a privileged place to develop a scientific culture which adds meaning to the acquired knowledge.

\end{document}